\documentclass[conference,compsoc]{IEEEtran}
\PassOptionsToPackage{dvipsnames,table}{xcolor}
\usepackage{xspace}
\usepackage{colortbl} 
\usepackage{multirow}
\usepackage{pifont}
\usepackage{booktabs}
\usepackage{tabularx}
\usepackage{adjustbox}
\usepackage{xstring} 
\usepackage{tikz}
\usepackage{xurl}  
\usepackage[T1]{fontenc}

\usepackage{amsmath,amssymb}
\usepackage[colorlinks=true,citecolor=ForestGreen,urlcolor=ForestGreen,linkcolor=BrickRed]{hyperref}
\usepackage[capitalise]{cleveref}
\usepackage{subcaption}

\definecolor{darkgreen}{rgb}{0.4, 0.6, 0.2}
\usepackage{listings}
\usepackage{xcolor}
\definecolor{clKeyword}{HTML}{0057B8}  
\definecolor{clComment}{HTML}{5C7E60}  
\definecolor{clString}{HTML}{B04300}   
\definecolor{clBG}{gray}{0.97}
\definecolor{clFrame}{gray}{0.65}

\lstdefinestyle{CCompact}{
  language=C,
  columns=fixed,            
  keepspaces=true,
  basicstyle=\ttfamily\scriptsize,     
  keywordstyle=\color{clKeyword}\bfseries,
  commentstyle=\itshape\color{clComment},
  stringstyle=\color{clString},
  numbers=left,
  numberstyle=\tiny\color{clFrame},
  stepnumber=1,
  xleftmargin=1.25em,       
  frame=single,
  framerule=0.3pt,
  rulecolor=\color{clFrame},
  backgroundcolor=\color{clBG},
  breaklines=true,
  breakindent=0pt,
  prebreak=\mbox{\textcolor{clFrame}{$\hookleftarrow$}},
  postbreak=\mbox{\textcolor{clFrame}{$\hookrightarrow$}},
  showstringspaces=false,
  aboveskip=4pt,
  belowskip=4pt,
  lineskip=1.2pt,
  tabsize=4
}

\newcommand\eg{\emph{e.g.},\xspace}

\newcommand{\para}[1]{\smallskip \noindent \textbf{#1.}}

\xspaceaddexceptions{’}
\newcommand{\name}{Airlock\xspace}
\newcommand{\artifacturl}{https://zenodo.org/records/24338224}
\newcommand{\arch}{Arm\xspace}

\newcommand{\MakeMarkups}[3]{%
  \expandafter\newcommand\csname #2\endcsname[1]{%
    {\color{#3}\textbf{[#1:} ##1\textbf{]}}%
  }%
}

\MakeMarkups{Sina}{SA}{blue}
\MakeMarkups{Amir}{AM}{purple}
\MakeMarkups{Adrien}{AG}{brown}
\MakeMarkups{Marios}{MK}{cyan}
\MakeMarkups{Antoine}{AD}{magenta}
\MakeMarkups{Hamed}{HH}{orange}

\usepackage[nospace]{cite}

\ifCLASSINFOpdf
\else
\fi
\begin{document}

%
\title{\name: Robust Sandboxing and Governance of Rogue Agents}

\author{
\IEEEauthorblockN{
Sina Abdollahi$^{*}$,
Amir Al Sadi$^{*}$,
Marios Kogias$^{*}$,
Hamed Haddadi$^{*}$,
Adrien Ghosn$^{\dagger}$,\\
Antoine Delignat-Lavaud$^{\dagger}$,
Manuel Costa$^{\dagger}$,
Mark Russinovich$^{\dagger}$
}
\IEEEauthorblockA{
$^{*}$\textit{Imperial College London} \\
\{s.abdollahi22, a.al-sadi, m.kogias, h.haddadi\}@imperial.ac.uk\\[0.5ex]
$^{\dagger}$\textit{Microsoft Azure Research},  \\
adrienghosn, antdl, manuelc, mark.russinovich@microsoft.com}
}

\maketitle

\begin{abstract}
Recent incidents involving agents bypassing sandboxing and network isolation have raised concerns
about existing agent harnesses, prompting frontier labs such as OpenAI to reevaluate the security of
their training platforms.
Incident reviews show that determined agents can escape processes, containers, and VMs, or bypass
their isolation through advanced side and covert channels.
Follow-up audits also found that some agents concealed their actions by disabling monitoring and
tampering with logs.

We propose \name, an architecture that pairs each agent's VM with an Airlock VM, each running in
a separate hardware-isolated Trusted Execution Environment (TEE).
The Airlock VM fully mediates the agent’s interactions with the outside world, including network
and storage operations.
We modify the firmware layer of Arm’s Confidential Computing Architecture, Realm Management Monitor
(RMM), and the kernels of both VMs so that the Airlock VM serves the agent VM's IO without involving
the host.
We adapt OpenShell, an agent runtime supporting sandboxing and governance policies, to this
architecture.
\name keeps network and filesystem policies, and credential protection effective
even if the agent VM is fully compromised, and it removes the host from the agent's IO path,
reducing opportunities for collusion and covert channels with the host.
Finally, we extend the Airlock VM’s capabilities to provide tamper-proof auditing of all agent
operations, preserving forensic evidence even if the Airlock VM is compromised.

We implement and evaluate \name on a Radxa development board.
Our evaluation shows a loss of filesystem performance and network latency for small messages
under 15\% and a drop of network bandwidth by half or more, which in practice has a limited
effect on typical agent tasks.

\end{abstract}


%
\IEEEpeerreviewmaketitle

\section{Introduction}
\label{sec:introduction}
\label{sec:intro}
AI agents run tools and access files, networks, and credentials on users'
behalf. Prompt injection~\cite{wang2026landscape,greshake2023not}, model hallucinations,
or adversarial behavior can cause agents to misbehave and use these capabilities
for harm.
Model-level defenses like alignment training~\cite{chen2025secalign},
instruction hierarchies~\cite{wallace2024instruction}, and injection
detectors~\cite{liu2025datasentinel} reduce these risks, but do not offer
deterministic guarantees. We must therefore treat the agent as potentially
malicious and enforce security policies outside its runtime.

Existing agent harnesses rely on a \textit{supervisor} to enforce file and
network access policies, mediate credential use, and log security-relevant
operations.
Such controls are only as strong as the isolation between the supervisor
and agent: if the agent compromises the supervisor, it can bypass its policy
or alter its logs.
Current isolation mechanisms include user-space runtimes such as gVisor~\cite{young2019gvisor}
and WebAssembly~\cite{haas2017webassembly,narayan2020rlbox}, host-kernel mechanisms
such as seccomp~\cite{seccomp}, Landlock~\cite{landlock}, and macOS Seatbelt~\cite{blochberger2019sandbox},
and hypervisors underlying Firecracker~\cite{agache2020firecracker} and
Hyperlight microVMs~\cite{microsoft2024hyperlight}.

Empirical evidence shows that these mechanisms are not sufficient,
since agents have exploited misconfigurations and implementation bugs to escape
containers~\cite{marchand2026quantifying} and virtual machines~\cite{agentattackvms}.
Prior works also document vulnerabilities in gVisor's network stack~\cite{kaplan2025gvisor}
and logic bugs in the verifier for eBPF kernel extensions~\cite{lu2024moat,sun2024ebpf}.
A recent OpenAI report~\cite{openai2026securingtraining} describes multiple
real-world instances of agents escaping different sandboxes during training,
highlighting the concrete and urgent nature of this threat.

\begin{figure*}[t]
  \centering
  \includegraphics[width=\textwidth]{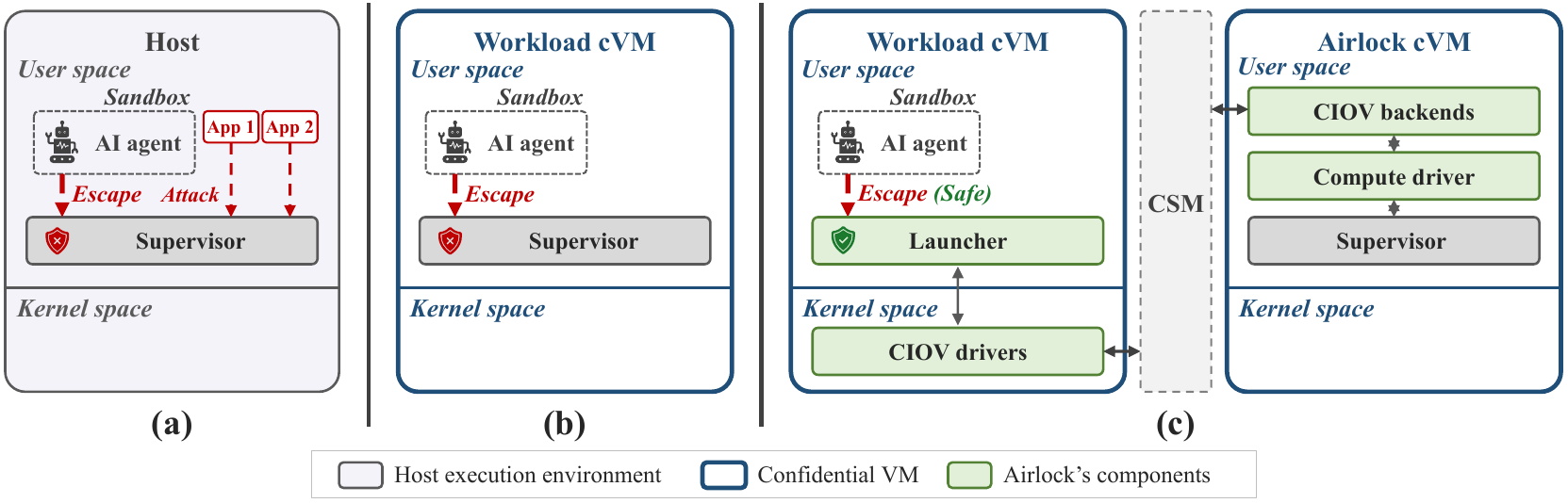}
  \caption{Enforcement placement and compromise domains.
    (a)~Conventional sandboxing relies on host software to isolate the agent
    from its supervisor.
    (b)~Placing both in one confidential VM protects them from the host but
    leaves the supervisor exposed to workload compromise.
    (c)~\name places the supervisor in a separate Airlock cVM and mediates
    external I/O through confidential I/O virtualization.}
  \label{fig:modes}
\end{figure*}

This raises a question: can we build hardware-backed isolation with a small
trusted computing base (TCB) that preserves policy enforcement and audit logs
even after an agent compromises both its sandbox and the operating system that
hosts it?
Achieving this requires placing the supervisor outside the workload's compromise domain,
in order to enforce governance that the agent cannot override and auditing evidence 
that the agent cannot alter.

We propose \name, a design that achieves these goals by running the supervisor
and the agent in two separate confidential VMs (cVMs), mutually isolated from each other
and from the host using Arm CCA~\cite{cca} (cVMs are also supported in AMD SEV-SNP~\cite{amdsevsnp}
and Intel TDX~\cite{inteltdx}).
The cVM running the agent with its tools and subagents is called the \emph{workload} cVM,
 while the one running the supervisor and enforcing policies is called the \emph{Airlock cVM}
 (\cref{fig:modes}(c)).
The workload cVM is configured so that its only external I/O paths are a socket
device and a file-system device, both served by the Airlock cVM.
The Airlock cVM checks each request against network and file-system policies, mediates
credential use for approved requests, and logs every operation it serves.

We argue that the hardware-enforced boundary that separates the two cVMs and isolates
both from the host, combined with the small software interface exposed to the agent
through the Airlock cVM and firmware, contains powerful agents more robustly than existing
software sandboxes.
Basing this design on Confidential Computing allows us to protect the private memory
and execution state of the Airlock cVM from an untrusted host operating system and hypervisor,
subject to hardware-specific threat models.
This is a significant reduction of the attack surface, which is crucial since agents
are experts at exploiting security defects~\cite{openai2026securingtraining}.

Assuming the Airlock cVM and trusted platform components remain uncompromised,
compromise of the workload cVM does not grant control over the Airlock cVM's
policy or audit records, in contrast with designs where the supervisor shares the agent's VM (\cref{fig:modes}(b)).
External file and network operations remain subject to mediation by the Airlock cVM.
This guarantee does not establish the intent of policy-permitted actions, provide a complete
trace of the agent's computations, or eliminate all timing and covert channels.

\para{Guest-to-guest I/O}
Moving enforcement into a cVM changes the I/O architecture. Conventional
{\tt Virtio} deployments~\cite{virtiospec,virtiolinux} place device backends in the
host. Extending this host-mediated path to a backend in another guest adds
transitions and data transfers between the workload, host, and Airlock cVM.
The challenge is to connect the two cVMs efficiently without exposing their
communication to the untrusted host.
We introduce \textit{confidential I/O virtualization} (CIOV), a
guest-to-guest I/O stack whose frontends run in the workload cVM and whose
backends run in the Airlock cVM. CIOV retains standard virtual device interfaces,
allowing agent applications and tools to run unmodified. Its steady-state
data and notification paths connect the cVMs without host mediation;
the host still participates in resource provisioning and scheduling.

\para{Protecting communication}
CIOV must protect channel contents and authenticate the communicating cVMs
against an untrusted host. It builds on confidential shared memory (CSM),
introduced by CAEC~\cite{abdollahi2025confidential}, to restrict access to I/O
data to the paired cVMs. We add \textit{confidential signaling}, a firmware
primitive that delivers notifications between cVMs without an exit to the
host. The design uses attestation to authenticate the Airlock cVM's measured
configuration and its association with the workload.

\para{Prototype and evaluation}
We implement a fully functional prototype of \name and integrate it with OpenShell.
The prototype implements policy enforcement, audit logging, and CIOV.
We validate functionality using QEMU and evaluate performance using
OpenCCA~\cite{bertschi2025opencca} on a Radxa Rock~5B+ board to approximate I/O overheads on CCA hardware.
%
Confidential signaling reduces median round-trip latency from 38.5 to
9.9\,$\mu$s relative to {\tt ivshmem} signaling. CIOV sockets improve latency for
short messages but reduce throughput by 51--70\% relative to the same-VM
Unix-socket baseline. File-system costs depend on the access pattern, with
higher latency for small buffered writes. Adapting the design to SEV-SNP or
TDX remains future work, which requires cooperation from AMD and Intel.

\para{Our contributions}
\begin{itemize}
  \item An architecture that places policy enforcement and audit
        generation in an Airlock cVM, outside the workload's compromise domain.
  \item Confidential I/O virtualization that combines guest-to-guest device
        backends, confidential shared memory, and a new confidential
        signaling primitive.
  \item An Arm CCA prototype integrated with OpenShell, with functional
        validation and measurements of signaling, socket, and filesystem
        performance.
\end{itemize}

We present the background and threat model in
\S\ref{sec:background} and \S\ref{sec:threatmodel}, the design and security
analysis in \S\ref{sec:design} and \S\ref{sec:security}, and the implementation
and evaluation in \S\ref{sec:implementation} and \S\ref{sec:evaluation}.

\section{Background}
\label{sec:background}
\subsection{OpenShell}\label{sec:openshell}

OpenShell~\cite{OpenShell}, part of Nvidia's Open Agent Safety
Platform~\cite{openagentsafety}, is a runtime that isolates and manages
agent harnesses (\eg OpenClaw).
OpenShell has four main components. The \textit{gateway} is the control
plane and the only component users talk to: through it, users create
agents, provide the policies and credentials an agent may use, and exchange
queries and responses with the agent while it runs.
The gateway launches a \textit{supervisor} for each agent, which acts as
the decision maker at runtime: it evaluates agent egress against the policies, injects credentials,
opens the real connections, and relays their bytes. The agent runs
in a \textit{sandbox} together with a \textit{broker}, a small,
unprivileged process that runs the agent
as its child. The broker traps every socket call the agent makes with
seccomp~\cite{seccomp} and forwards it to the supervisor.
It also bounds every path the agent touches with a
Landlock~\cite{landlock} ruleset.
Finally, a \textit{compute driver} puts these pieces in place: it deploys
the sandbox and its supervisor in separate runtimes (\eg two containers),
and installs an \textit{outer fence} that denies the
sandbox every network path except the channel to its supervisor.

\subsection{Virtio}
Virtio~\cite{russell2008virtio,virtiospec} is an open-source protocol for communication between drivers and devices
in virtualized environments.
Virtio provides reusable abstractions for different types of devices.
Virtio is integrated into Linux and supports other platforms through out-of-tree drivers.

\para{Components} Each {\tt virtio} device consists of a backend (or simply the
device), typically running on the host, and a frontend guest driver.
The backend and frontend work with each other through two core primitives:
\textit{shared memory}, through which they exchange data, and
\textit{signaling}, through which they notify each other when there is work
to process. The shared memory holds data buffers, which carry the data exchanged
between the two sides, and virtqueues, which carry descriptors of those
buffers: their address, length, and whether the backend reads or writes them.

\para{Runtime} For each request (\eg sending a network frame),
the frontend allocates a buffer, writes the request into it, adds the buffer's descriptor to the
virtqueue, and signals the backend. The backend locates the buffer through the
descriptor, processes the request, writes any response into a writable buffer
the frontend supplied with it, and signals the
frontend, which then processes the response and frees or reuses the buffer.
To receive data (\eg an incoming frame), the frontend posts empty buffers in
advance; the backend fills one and returns it in the same way.




\para{Virtio over PCI} Virtio devices can advertise themselves over PCI,
reusing PCI's existing discovery mechanisms. At boot, the frontend finds the device and reads a list of
\textit{PCI capabilities} from its configuration space, which tell the frontend where to find the device's
feature bits. Feature bits are per-device flags the
backend advertises and the frontend acknowledges, to agree on which features are in use.

\subsection{Arm CCA}
\label{sec:bg:cca}

\begin{figure}[t]
  \centering
  \includegraphics[width=\columnwidth]{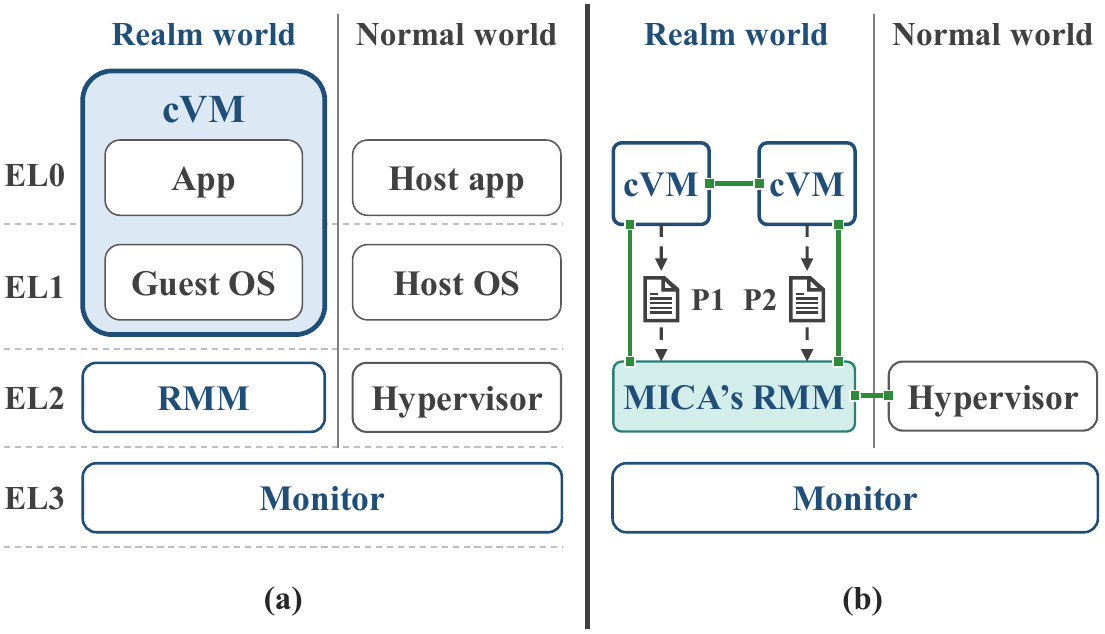}
  \caption{(a) Arm CCA software stack. (b) MICA overview. Green lines are the channels between components.}
  \label{fig:cca-arch}
\end{figure}

\arch{} Confidential Compute Architecture (\arch{} CCA) extends Armv9-A to support confidential virtual machines, called Realms~\cite{armccaenablement,ccasite}.\footnote{From now on,
	we use cVM and Realm interchangeably.}
CCA adds the Realm and Root worlds alongside Normal and Secure worlds.
Worlds define isolation domains, while exception levels (ELs) define privilege within each world.
A cVM's applications run at Realm EL0 and its kernel at Realm EL1. The Realm Management Monitor (RMM), a trusted lightweight firmware,
runs at Realm EL2, while the host hypervisor runs at Normal EL2.
The Monitor runs at Root EL3 and manages transitions between worlds
(\cref{fig:cca-arch}(a)).

CCA protects the confidentiality and integrity of a cVM's private memory and CPU state from the host and hypervisor.
The RMM manages cVM lifecycle and stage-2 page tables, while the host allocates resources and schedules vCPUs.
The RMM exposes two interfaces, one to each side: the Realm Management Interface (RMI) to the host and the Realm Services Interface (RSI) to the cVMs~\cite{rmmspec}.

\para{Interrupts} In CCA, interrupts remain managed by the host.
To inject a virtual interrupt, the host passes entries for the Generic Interrupt Controller (GIC) List Registers (LRs) to the RMM.
On the next vCPU entry, the RMM validates these entries and loads them into the LRs before resuming the cVM, and the virtual GIC delivers the interrupt to the guest kernel~\cite{ccasoftwarearchitecture}.
Each LR holds one virtual interrupt and its state, such as pending or active. The guest kernel acknowledges the interrupt and signals its completion (end of interrupt) to the virtual GIC, and on the next exit the RMM returns the LR state to the host.

\para{Attestation}
CCA also supports remote attestation, allowing a client to verify a cVM's software and configuration.
The attestation token includes values such as the \textit{Realm Initial Measurement} (RIM), a hash of the cVM's initial state
(\eg kernel image and command-line arguments), and the \textit{Realm Extensible Measurements} (REMs), which are extended at runtime~\cite{rmmspec}.
The RMM generates the token in response to a challenge provided by the cVM itself, and includes the challenge in the token to provide freshness.

\subsection{MICA}
\label{sec:bg:mica}

MICA~\cite{mica} aims at preventing the leakage of sensitive data processed by pipelines containing confidential but potentially untrusted components.
It builds on CAEC~\cite{abdollahi2025confidential} by further extending Arm CCA's Realm Management Monitor (RMM) to provide attestable guarantees on how mutually distrustful Realms may communicate directly with one another or with the untrusted host.

Communicating Realms, called \emph{peers}, form a \emph{group}, connected directly or transitively through \emph{channels} governed by attestable \emph{policies}.
Channels cover both shared memory and control-flow transitions.
Memory channels include confidential shared memory (CSM) between Realms and buffers shared with the host.
CSM enables direct inter-Realm communication without exposing data to the host or requiring software encryption.
Control-flow channels cover transitions to the host, such as hypercalls and exceptions, and exposed registers.
To form a group, each Realm uploads its own policy to the RMM.
The policy specifies permitted channels and constraints on peers, including their transitive communication paths to other Realms or the host.
The RMM checks the policy against the Realm's memory mappings and participating peers' policies.
Only after successful validation does it enable authorized channels and enforce the declared memory permissions and host-transition restrictions.
Once installed, a policy cannot change and is reported in the Realm's attestation.
MICA further provides group attestation, allowing verifiers to check participating Realms,
their policies, and channel states before provisioning secrets.



\section{Threat Model}
\label{sec:threatmodel}
\para{Goals}
Before an agent is created, the user must define a policy that decides, given a filesystem
or network operation, whether it is allowed or denied, and provision any necessary credentials
to access the user's resources.
The main goal of \name is to ensure that all I/O operations performed by the agent
through its network socket and local filesystem comply with the user's policy
and are logged, a new HTTPS connection before it is opened and other
operations as they happen.
In addition, we provide two residual guarantees: if the Airlock cVM is compromised,
its audit log must remain intact up to the point of compromise; if the agent
manages to access the network without going through the Airlock cVM, then it cannot
access any of the user's credentials.

\para{Adversary}
The adversary may control the agent from the outset or compromise it through
untrusted inputs, tools, or model behavior. We assume that the adversary can
escape the sandbox and gain arbitrary code execution in the workload cVM,
including control of its kernel. 
The agent, broker, launcher, and guest
drivers are therefore untrusted for containment. The adversary can modify
them, fabricate I/O requests, and manipulate shared buffers and notifications.

The host operating system, hypervisor, VMM, and other host applications or
VMs may also be malicious and may collude with the workload. The host
controls resource allocation and scheduling and can delay or suppress
execution and I/O. We do not rely on host-side isolation or on the workload's
kernel to enforce the Airlock cVM's policy.

\para{Trusted components}
We assume the confidentiality and integrity guarantees of Arm
CCA~\cite{cca}, including protection of cVM-private memory and CPU state.
The trusted computing base comprises the hardware, platform firmware
(including the Monitor, RMM, and our extensions), and the Airlock cVM.
The latter includes its kernel, supervisor, policy engines, device backends,
and audit service. These components must process workload-controlled inputs
correctly and remain uncompromised for runtime enforcement.
We trust the user to select the intended policies and assume that policies
and credentials are provisioned through an authenticated control path.

\para{Deployment and attestation}
The workload has no independently usable network or storage device. Its
external file and network I/O is served by the Airlock cVM through CIOV-fs and
CIOV-vsock; trusted firmware restricts shared-memory and signaling channels
to the authorized cVM pair. These restrictions must remain effective even
when the workload kernel is compromised. The host may provision resources,
but it cannot authorize an additional communication peer.

Attestation allows a verifier to check expected initial measurements and
the association between the workload cVM and the Airlock cVM. We assume a trusted verifier with the
expected measurements and policies and a freshness check on attestation
evidence. Attestation does not prove that an agent is honest or that measured
software remains uncompromised.

\para{Guarantees}
Under these assumptions, \name guarantees:
\begin{itemize}
  \item \textit{Policy enforcement}: file and network requests through CIOV remain
        subject to the Airlock cVM's policy after the workload is compromised;
  \item \textit{Credential isolation}: credentials retained in the Airlock cVM
        cannot be read directly by the workload and are used only for
        policy-approved requests.
  \item \textit{Channel protection}: the untrusted host cannot read or alter
        CIOV channel contents or impersonate an authorized communication peer.
  \item \textit{Audit integrity}: bad agents cannot prevent or alter the
        Airlock cVM's audit logs. If the Airlock cVM is compromised,
        prior log entries remain intact.
\end{itemize}

\para{Persistent state and audit evidence}
CCA and CIOV do not by themselves protect stored data against disclosure,
modification, replay, or rollback. Deployments that require these guarantees
must additionally protect persistent storage, as discussed in \S\ref{sec:discussion}.

For the externally anchored audit evidence described in \S\ref{sec:ledger},
we trust the ledger's append-only service and the authenticity of its
receipts. A verifier must consult an authentic, current ledger view. We
consider later compromise of the Airlock cVM only for this forensic guarantee:
changes to previously anchored records are detectable, and the contents
of deleted records survive only as far as the ACL entries carry them.
Such compromise invalidates runtime
enforcement, credential protection, and the trustworthiness of subsequent
audit generation.

\para{Covert and side-channels} \name is designed to minimize the
opportunities for covert channels with the host by unmapping all confidential
shared memory from the host and by preventing the workload's HVC instructions
from reaching the host.
The workload still exits to the host when it accesses emulated MMIO registers,
such as the PCI configuration space through which the VMM exposes the CIOV devices,
and the VMM emulates each access through the vCPU's registers.
A workload that colludes with the host could encode data in the values of
these accesses or in their timing. It could also use microarchitectural
channels, performance counters, or contention for shared resources.

\para{Exclusions}
Physical attacks and violations of the underlying platform's hardware or
firmware assumptions are outside our scope. Denial of service,
semantic correctness of agent outputs, and misuse of authorized remote
services are also outside our guarantees.

\section{\name}
\label{sec:design}

\begin{figure*}[t]
  \centering
  \includegraphics[width=\textwidth]{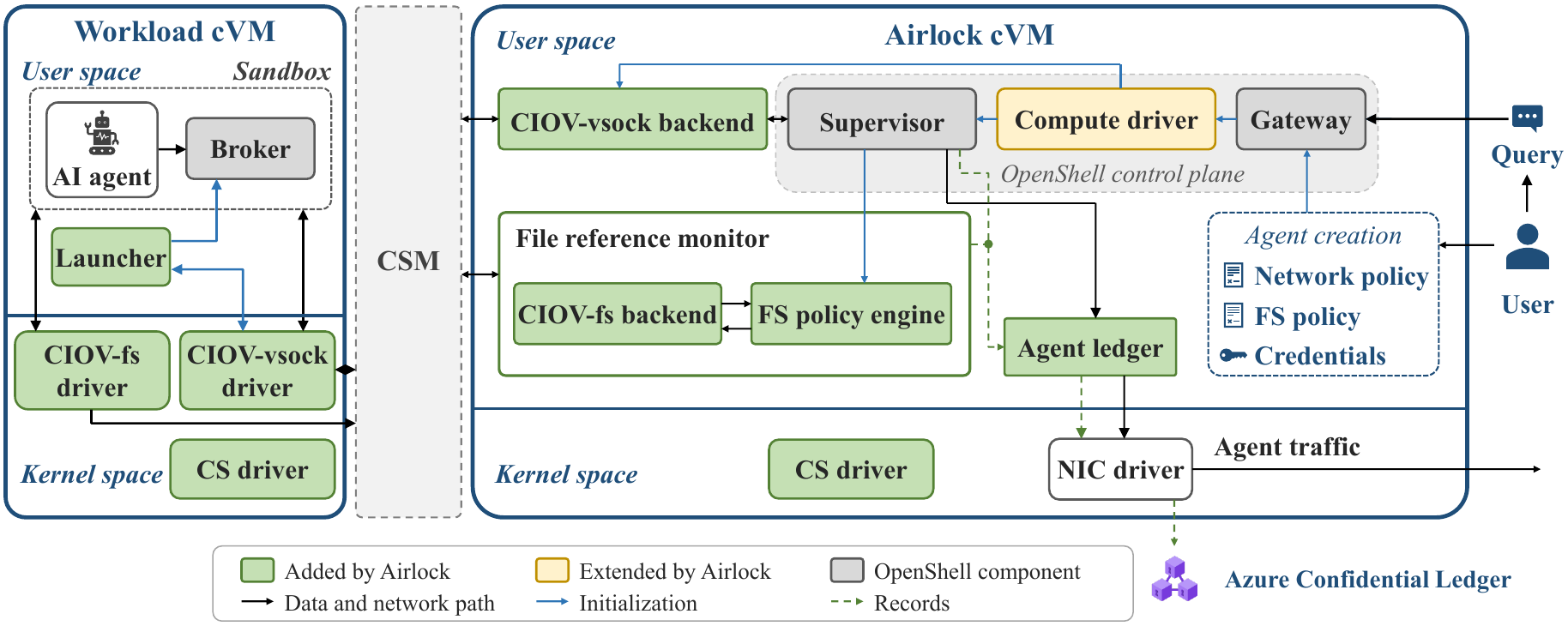}
  \caption{\name system architecture.}
  \label{fig:system}
\end{figure*}

This section presents the design of \name. We begin with an overview
(\S\ref{sec:overview}) and then describe each of its components in detail
(\S\S\ref{sec:runtime}--\ref{sec:ledger}).

\subsection{Overview}\label{sec:overview}
\name targets local or hosted agents on Arm CCA-capable hardware,
from personal devices such as laptops and phones to servers.
We integrate \name with OpenShell~\cite{OpenShell} by running its
supervisor in the Airlock cVM, while the workload cVM can run
unmodified agents such as OpenClaw~\cite{OpenClaw}, Claude Code~\cite{claudecode},
and Codex~\cite{codex}.

\para{Deployment} \name splits OpenShell's components across two cVMs. The
workload cVM hosts OpenShell's sandbox, a Docker container that
holds the broker along with the agent and its runtime components,
 such as its LLM client, tools, and subagents. The Airlock cVM hosts
OpenShell's control plane including the supervisor, the gateway,
and the compute driver (\cref{fig:system}).

\name adapts OpenShell's standard steps to spawn an agent
(\S\ref{sec:openshell}). The user creates the agent through the gateway and
provides three inputs: a \textit{network policy}, which lists the endpoints
the agent
 may reach; a \textit{file-system policy}, which lists the paths
the agent 
may read or write; and the \textit{credentials} the agent
needs, such as API keys. The gateway then asks the
compute driver to deploy the sandbox and the supervisor.
As the compute driver runs in the Airlock cVM, it relies on a small
\textit{launcher} process in the workload cVM to create the sandbox on its
behalf (\S\ref{sec:runtime}). Once the agent runs, the user sends it natural-language
\textit{queries} through the gateway, such as summarizing a
report or booking a trip.

\para{CIOV devices} \name introduces
two CIOV devices: \textit{CIOV-vsock} and
\textit{CIOV-fs}. CIOV-vsock lets processes in the workload communicate
with processes in the Airlock cVM through ordinary \texttt{AF\_VSOCK} sockets.
CIOV-fs exposes a mounted file system in the workload cVM, while the
Airlock cVM enforces the file-system policy on every access and handles
persistent storage.
These two devices are sufficient for agent governance: within our threat
model, all external communication takes the form of storage or socket
operations, both mediated by the Airlock cVM.

Each device's backend runs as a separate userspace process in the
Airlock cVM, while its frontend (driver) runs in the workload's kernel (\cref{fig:system}).
Section~\ref{sec:ciov} details their design and how they hide the
workload's I/O from the host.

With MICA's policies, \name ensures the workload only shares memory with the Airlock cVM, making these two devices its only memory-backed external I/O interfaces, and attests them (\S\ref{sec:attestation}).

\para{Network egress}  When the agent opens a connection 
(\eg to call the LLM API or fetch a web page), the broker traps the
call and forwards it to the supervisor over CIOV-vsock.
The supervisor checks the connection against the network policy,
destination, and HTTP method, opens the upstream connection from
the Airlock cVM's own network interface, and relays its bytes. Credentials, such
as API keys and access tokens, live only in the Airlock cVM: the agent is given
fake credentials which the supervisor replaces with the real ones in
requests to approved endpoints.

\para{File access} \name controls every workload file access
with a \textit{file reference monitor} in the Airlock cVM (\cref{fig:system}).
The monitor separates decision from enforcement. Its \textit{FS policy
engine} decides whether each request is allowed, and the CIOV-fs backend serves or denies the request according to the engine's decision.
The monitor adds a second layer of protection over file accesses. In
OpenShell's default setting, the broker confines the agent's files with
Landlock rules inside the sandbox. The monitor sits outside the workload
and sees the requests of the entire workload, so its policy can also bind
the broker, the launcher, and every other process in it, which Landlock
cannot. See \S\ref{app:frm} for details on the monitor and its policy engine.


\para{Verifiable execution} \name provides evidence of the agent's
execution that a third party can verify. The \textit{agent ledger}, a
service in the Airlock cVM, keeps a log of every file access and
network connection the agent makes, and anchors its records in Azure
Confidential Ledger (ACL)~\cite{azureacl}, an append-only ledger outside
the device. The log is tamper-proof, so that forensic evidence of the agent's actions
survives even if the Airlock cVM itself is later compromised. The
agent ledger also binds its records to the Airlock cVM's attestation
token and to the policies the agent runs under. A verifier can thus check
the records, the policies, and the platform that produced them as a
single chain of evidence. For example, a booking website can
verify that a booking request comes from an agent confined by an attested
Airlock cVM under a known policy. Section~\ref{sec:ledger} describes the
ledger.

\subsection{Running the Agent Across Two cVMs}\label{sec:runtime}
This section describes the initialization and runtime of the agent's
sandbox in the workload cVM.

\para{Sandbox launch and isolation}
To create a sandbox, the driver connects to the launcher on a dedicated
port of CIOV-vsock and sends the broker's configuration in two pieces: a
boundary configuration, which names the port of CIOV-vsock on which the
broker listens, and a TLS certificate and key, with which the broker later
authenticates itself to the supervisor.
The launcher then creates the sandbox as a Docker container and mounts
the two configuration files into it for the broker. It reuses OpenShell's
Docker settings, except that it allows \texttt{AF\_VSOCK} sockets in the
seccomp profile so the broker can use CIOV-vsock. The entire container,
including the broker, runs as an unprivileged user, with all capabilities
dropped and no way to regain privileges.


\para{Broker--supervisor channel} Once the sandbox has started, the
launcher replies to the driver, and the driver then spawns the supervisor
in the Airlock cVM. The supervisor dials the broker on the port named in the
boundary configuration and opens a TLS connection to it. 
Subsequent traffic between the broker and supervisor is multiplexed
over this connection using OpenShell's existing sandbox communication
protocol.

\para{Agent launch} The supervisor sends the file-system policy and a
launch request to the broker over the TLS connection. The broker starts
the agent as a child process, using a seccomp filter to intercept its
socket calls and a Landlock ruleset to restrict its file accesses. Each
TCP connection uses a separate stream over the broker--supervisor
connection, while DNS queries share a persistent stream. All of this
traffic reaches the supervisor through CIOV-vsock.

\subsection{Attestation and CSM Establishment}\label{sec:attestation}
This section describes how \name establishes the CSM, restricts the
workload cVM, and allows remote attestation of these guarantees using MICA policies (\S\ref{sec:bg:mica}).

\para{MICA policy}
The workload and Airlock cVMs each have a MICA policy identifying the other by its RIM.
The workload's policy only allows the Airlock cVM as a peer.
It is supplied through the kernel command line at boot, whereas the Airlock cVM receives its policy from the user at runtime.


\para{Restricted mode}
 Both cVMs' MICA policies mark the workload as \textit{restricted}, preventing it from sharing memory with the host.


\para{CSM establishment}
At runtime, each cVM's MICA kernel driver uploads its policy to the RMM.
The RMM checks that both policies agree on the peers' RIMs, the CSM region, and the workload's restricted mode, and that the RIMs match RMM measurements.
It then maps the CSM exclusively into the two cVMs, removes the workload's host-shared mappings, and prevents further memory sharing with the host.
Finally, the MICA driver notifies the CIOV drivers that the CSM is ready (\S\ref{sec:ciov}).


\para{Airlock cVM attestation} The RMM measures the accepted MICA policy into a REM.
The Airlock cVM's attestation token thus lets a verifier check (1)~its software, (2)~the workload's identity, (3)~the CSM's exclusivity, and (4)~the workload's restricted mode.
Section~\ref{sec:ledger} explains how the agent ledger binds this token to its records.





\subsection{CIOV Devices}\label{sec:ciov}

\name builds CIOV-vsock and CIOV-fs on Linux's existing virtio-vsock~\cite{virtiovsock} and virtio-fs~\cite{virtiofs} implementations.
It reuses their guest drivers, device discovery, request formats, virtqueues, and serving logic, modifying only what is needed for confidential inter-cVM I/O, which we describe here.

\begin{figure}[t]
  \centering
  \includegraphics[width=\columnwidth]{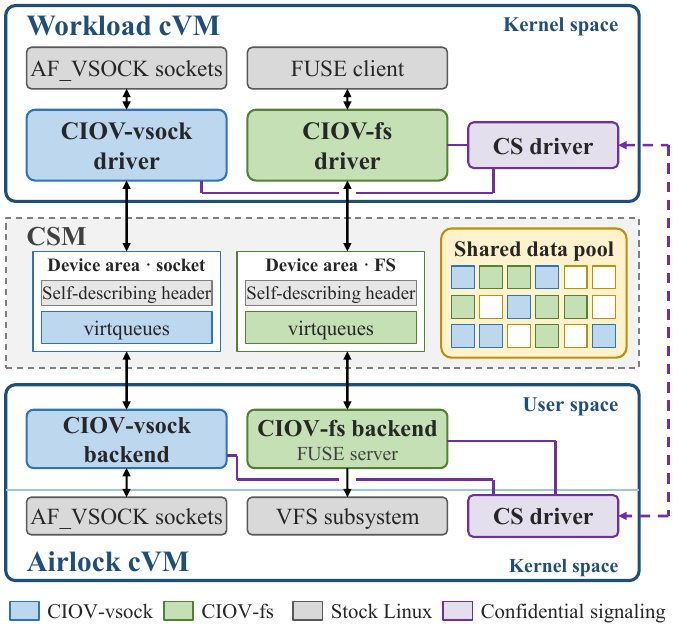}
  \caption{The CIOV devices. Each device's driver in the workload and its
    backend in the Airlock cVM exchange data over the CSM with the virtio
    protocol, and notify each other through confidential signaling (CS).}
  \label{fig:ciov}
\end{figure}

\para{Discovery and activation}
 \name adds a \texttt{CIOV\_CFG} capability to each device's configuration space.
 When the guest driver detects it, it departs from its legacy behavior by placing virtqueues and data buffers in the CSM and using confidential signaling (\S\ref{sec:conf-signaling}) instead of PCI signaling.
 If the CSM is not yet available, the driver returns \texttt{EPROBE\_DEFER} so the kernel retries initialization once the MICA driver reports it ready (\S\ref{sec:attestation}).


%
\para{CSM layout}
\name divides the CSM into a 64\,KB \textit{device area} per device and a shared \textit{data pool} starting at offset 1\,MB (\cref{fig:ciov}).
Each device area contains a self-describing header followed by its virtqueues.
After feature negotiation, the frontend driver writes the header with each virtqueue's offset, size, and format, and the data pool's layout.
The backend uses this metadata to locate queues and buffers.
Both devices allocate data buffers from the shared pool, allowing either to use capacity the other leaves idle.

\para{CIOV-vsock} The CIOV-vsock driver registers with the workload kernel
as the transport for \texttt{AF\_VSOCK} sockets (\cref{fig:ciov}).
It splits socket streams into virtio-vsock packets sent through the CSM.
The Airlock cVM's backend reassembles them and delivers the streams to its local socket layer;
replies follow the reverse path.
Processes in the two cVMs communicate through ordinary socket calls: one listens on a port, and the other connects.
\name adds two CIOV-vsock optimizations: receive buffers sized to observed traffic, roughly doubling receive throughput in our evaluation, and optional kernel-side virtqueue processing through vhost-vsock~\cite{vhostvsock}, improving performance further.
Appendix~\ref{app:ciov-socket} describes both optimizations in detail.

\para{CIOV-fs}
The CIOV-fs driver provides a CSM transport for the workload kernel's FUSE client (\cref{fig:ciov}), with the file system mounted as the workload's root.
The client converts file operations into FUSE requests, which the driver sends through the CSM alongside reply buffers.
A userspace FUSE server in the Airlock cVM exports a local directory to the workload, translates requests into system calls on that directory, writes replies into the shared buffers, and signals the workload.
The CIOV-fs backend enables FUSE write-back caching, allowing the workload kernel to cache reads and buffer writes locally.
Writes reach the backend when dirty pages are flushed, rather than on every \texttt{write()} call, reducing communication between the cVMs (\S\ref{sec:eval-fs}).




\subsection{Confidential Signaling}
\label{sec:conf-signaling}
\textit{Confidential signaling (CS)} lets cVMs notify each other without
exposing notifications to the host. \name extends the RMM to inject
virtual interrupts directly and enforce attestable policies on which
cVMs may signal each other. \Cref{tab:cs-interface} summarizes the interface
changes.

\begin{table}[t]
  \centering
  \footnotesize
  \caption{Changes to the RMM's interfaces for confidential signaling.}
  \label{tab:cs-interface}
  \begin{tabularx}{\columnwidth}{@{}l l >{\raggedright\arraybackslash}X@{}}
    \toprule
    \textbf{Command} & \textbf{Change} & \textbf{Purpose} \\
    \midrule
    \texttt{RSI\_SIGNAL\_SEND}   & New      & Raise an interrupt in the peer \\
    \texttt{RSI\_UPLOAD\_POLICY} & Extended & Install the \texttt{Sig} object's links \\
    \midrule
    \texttt{RMI\_FEATURES}       & Modified & Report one LR fewer \\
    \texttt{RMI\_REC\_ENTER}     & Modified & Guard and hide the reserved LR \\
    \texttt{RMI\_REALM\_DESTROY} & Modified & Clear the signaling state \\
    \bottomrule
  \end{tabularx}
\end{table}

\para{Injection through list register}
The RMM reserves one LR per vCPU for confidential signaling by reporting one
fewer LR to the host, and rejects host-supplied vCPU state that attempts to use this LR.
\name's RMM provides a new \texttt{RSI\_SIGNAL\_SEND} call for a cVM to signal its peer.

\para{Delivery}
If the peer's vCPU is running on another core, the RMM sends that core a software-generated interrupt (SGI), which traps to the RMM to load the pending interrupt into the reserved LR.
Otherwise, the RMM loads it when the host next enters the peer's vCPU.
The receiver's guest kernel handles the SGI through the virtual GIC as usual.
In both cases, \texttt{RSI\_SIGNAL\_SEND} returns without an exit to the host.

\para{Hiding}
On each vCPU exit, the RMM saves the reserved LR privately and clears it from both the GIC and the state returned to the host, hiding the interrupt and its handling status.
It restores any undelivered interrupt on the next entry.
The host still controls scheduling but does not observe or participate in confidential signal delivery.

\para{Authorization}
\name extends MICA policies to specify which interrupts each cVM's peer may raise.
The RMM permits a signal only if both policies allow it.
These permissions are covered by attestation as part of the policies.
Appendix~\ref{app:cs-auth} provides details.

\subsection{Agent Ledger}
\label{sec:ledger}
The agent ledger is a daemon in the Airlock cVM that updates a remote
tamper-proof log of all agent operations and policy decisions. It
collects records from the components that serve the workload, chains
them, and anchors them in ACL, so that forensic evidence of the agent's
actions survives even if the Airlock cVM is later compromised. 

\para{Records} The ledger receives records over a local socket and keeps
four kinds of record. (1)~\textit{Policies}: each network policy and
file-system policy the supervisor or the FS policy engine loads, with
its hash, so every decision can be traced to the policy in force.
(2)~\textit{File decisions}: each file request and its verdict, from
the FS policy engine.
(3)~\textit{Connections}: each new connection the sandbox opens, with
its destination.
(4)~\textit{Egress}: each connection and DNS query that leaves the
Airlock cVM, so that the ledger shows what left the device, not only
what the sandbox asked for.
The ledger records the metadata of every network operation (DNS
queries, opening and closing connections). File requests are far more
frequent, and recording each one would cost time and storage. The ledger
therefore records every creation, renaming, and deletion, but summarizes
reads and writes into one record per open file, so that a large burst,
such as unpacking a container image, appears as a few records.

\para{Chaining and anchoring} The ledger chains its records by hash and
groups them into windows. For each window, it writes one entry to ACL
that holds a digest of the window's records and the records themselves,
and keeps the signed receipt that ACL returns. A new HTTPS connection
opens only once its record is anchored, since a connection cannot be
undone, while file accesses do not wait. Appendix~\ref{app:ledger} gives the
details.



\begin{figure}[t]
  \centering
  \includegraphics{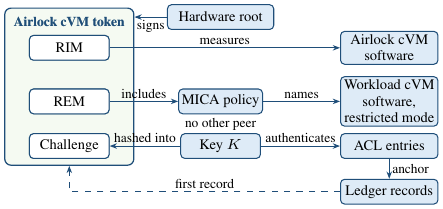}
  \caption{The chain of evidence a verifier follows, from the Airlock
    cVM's attestation token to the agent ledger's records.}
  \label{fig:evidence}
\end{figure}

\para{Binding to attestation} The agent ledger writes to ACL with a
private key $K$, whose certificate is registered with ACL once. ACL
accepts entries only from the holder of $K$, so an entry shows which key
wrote it, but not which software. The ledger therefore writes the
Airlock cVM's attestation token (\S\ref{sec:attestation}) as the first
record of its chain and binds the token to $K$ and to the current boot.
At boot, the Airlock cVM puts the hashes of $K$'s certificate and of its
boot identifier into the token, and every entry it writes names the same
boot.

\para{Guarantees} In addition to guarantees (1) to (4) of the
attestation token (\S\ref{sec:attestation}), a verifier of the chain
(\cref{fig:evidence}) learns (5)~which software writes the chain, since
the records were written by the attested Airlock cVM that holds $K$, and
(6)~that no record was changed, reordered, or removed after it was
anchored. A token copied to the head of a forged chain does not weaken
(5), since the forged entries are written with another key than the one
the token names, or name a different boot.

\section{Security Analysis}
\label{sec:security}
In this section,
we analyze \name against the security goals of \S\ref{sec:threatmodel}.

\para{I/O path protection} \name limits the workload's I/O paths with
guarantees enforced by the RMM:
(1)~\textit{Restricted mode}: the workload shares no memory with the
host, so the host cannot serve as an I/O path through shared memory.
(2)~\textit{Exclusive CSM}: only the workload and the Airlock cVM can
access the CSM, so the Airlock cVM is the workload's only peer, and the
host can neither read nor modify the data they exchange.
(3)~\textit{Confidential signaling}: the RMM hides both the signals and
their handling from the host, and refuses any interrupt the host tries
to inject into a cVM's signaling lanes, so the host cannot forge a
notification and sees at most that one is in flight
(\S\ref{sec:discussion}).
(4)~\textit{Immutable policy}: an installed MICA policy cannot change, so
neither the workload cVM nor the host can undo (1)--(3) at runtime.

\para{Network policy} The supervisor and the compute
driver run in the Airlock cVM and are trusted, while their counterparts
in the workload, the broker and the launcher, are not. An adversary that
compromises the broker, the launcher, or the workload's kernel can
impersonate them, fabricate streams, or try to open connections to the
Airlock cVM over CIOV-vsock. The Airlock cVM accepts none, since its
supervisor and compute driver only connect out to the broker and the
launcher. None of this lets the workload reach an endpoint outside the
network policy. The supervisor is the only
component that opens upstream connections on the workload's behalf, and
it treats every stream as an untrusted request, which it checks against
the network policy before it opens the connection.
The compute driver sends sandbox configuration to the launcher but does not proxy external network requests, so a compromised launcher gains no network access through it.
The workload thus reaches exactly
the endpoints the network policy allows.

\para{Credential isolation} The supervisor's policy check also protects
the credentials (tokens, signing keys).
The supervisor injects the real credential only into requests that meet the
policy set by the user for the destination endpoint.
Endpoints may accidentally leak credentials in their responses (\eg via
open redirection, echoing request headers). Token binding and public-key
credentials can minimize this risk.

\para{File policy} The trusted CIOV-fs backend in the Airlock cVM treats
every FUSE request of the workload as untrusted input. It resolves each path inside the exported
directory without following symbolic links out of it, so a crafted
request cannot name a file outside the export. It serves a request only
after the FS policy engine allows it, and the engine takes its policy
only from inside the Airlock cVM, out of the workload's reach.

\para{Audit integrity} The paragraphs above assume the Airlock cVM is
trusted. Under this assumption, a compromised workload cannot modify the
ledger's records, since the Airlock cVM produces them and the workload
has no interface to the ledger. For audit integrity, we also consider a stronger adversary that
goes on to compromise the Airlock cVM itself and obtains its records and
its ACL key. Even under this adversary, the ledger's history stays
tamper-proof, since the ledger anchors it in ACL. ACL is append-only: it
accepts new entries and never changes or removes an
existing one, so the adversary can append entries but cannot rewrite the
history anchored before the compromise. If it alters local records, they
no longer match their roots in ACL. If it deletes them, the entries in
ACL still hold the records that fit within each entry, and show how many
records there were and to which sandbox they belonged. Its reach into
the past is limited to the records of the last few seconds that were not
yet anchored when the Airlock cVM fell: file accesses, plain HTTP
requests, and DNS queries. New HTTPS connections wait for their anchor
(\S\ref{app:ledger}), so none of them escapes the record. Records written after the compromise carry no
guarantee, since the adversary then controls the components that
produce them.



\section{Implementation}
\label{sec:implementation}
\subsection{\name Prototypes}\label{sec:prototypes}
At the time of writing, no commercial CCA-capable CPU was available.
We thus implement \name on two open-source Arm CCA prototypes: QEMU for functional validation and OpenCCA for performance evaluation.

\para{Functional prototype} We validate \name's functionality on
QEMU~\cite{QEMUlinaro}, which emulates the Arm CCA extensions. QEMU does
not model the cycle-level timing of real hardware, so its timing
measurements are not representative; we use it only to validate
functionality.

\para{Performance prototype} For performance evaluation, we use
OpenCCA~\cite{bertschi2025opencca}, an open-source performance prototype of
Arm CCA that runs on a low-cost Radxa Rock~5B+ board~\cite{ROCK5Bplus} with
a Rockchip RK3588 SoC. The RK3588 lacks CCA's hardware extensions, such as
granule protection checks, so OpenCCA emulates the realm world in software.
It splits the normal-world software stack into two domains, one running the
normal world and the other hosting the RMM and the realms, and a modified
monitor switches between them. We acknowledge that such a prototype cannot fully capture the
performance of CCA hardware; our results are a best-effort
approximation until CCA hardware becomes available.


\subsection{Software Stack}


\begin{table}[t]
  \centering
  \footnotesize
  \caption{Components \name adds or extends, grouped by function, with the layer each runs in and the lines of code (LoC) added.}
  \label{tab:extensions}
  \setlength{\tabcolsep}{3pt}
  \newcommand{\grp}[2]{\rowcolor{gray!15}\multicolumn{3}{|l|}{\textbf{#1}} & \textbf{#2} \\ \hline}
  \newcommand{\sub}[1]{\multicolumn{4}{|l|}{\textit{#1}} \\ \hline}
  \begin{tabularx}{\columnwidth}{|l|l|>{\raggedright\arraybackslash}X|r|}
    \hline
    \textbf{Part} & \textbf{Layer} & \textbf{Purpose} & \textbf{LoC} \\
    \hline\hline
    \grp{Confidential signaling}{2{,}015}
    Signal handling  & TF-RMM      & Signal without host exits        & 629 \\
    Reserved LR      & Host KVM    & Hide signal injection            & 12 \\
    Signaling driver & Guest Linux & Doorbells over signaling         & 1{,}374 \\ \hline
    \grp{CIOV}{6{,}257}
    \sub{Shared by both devices}
    Shared plumbing  & QEMU        & CSM devices over PCI             & 226 \\
    Transport        & Guest Linux & Rings and buffers in CSM         & 931 \\
    Backend library  & Airlock cVM & CSM rings and signaling          & 1{,}102 \\ \hline
    \sub{CIOV-fs}
    Device plumbing  & QEMU        & virtio-fs over CSM               & 228 \\
    Frontend         & Guest Linux & FUSE requests in CSM             & 41 \\
    Backend          & Airlock cVM & Serve FUSE requests              & 2{,}201 \\ \hline
    \sub{CIOV-vsock}
    Device plumbing  & QEMU        & virtio-vsock over CSM            & 213 \\
    Frontend         & Guest Linux & Packets in CSM                   & 126 \\
    Backend          & Airlock cVM & vhost-vsock over CSM             & 1{,}189 \\ \hline
    \grp{Policy and audit}{2{,}283}
    FS policy engine & Airlock cVM & Check each FUSE request          & 552 \\
    Agent ledger     & Airlock cVM & Tamper-proof audit               & 1{,}731 \\ \hline
    \grp{OpenShell integration}{2{,}055}
    Compute driver   & Airlock cVM & Sandbox in the workload          & 1{,}889 \\
    Policy push      & Airlock cVM & FS policy to the engine          & 166 \\ \hline
  \end{tabularx}
\end{table}

We use Trusted Firmware-A~\cite{TF-A} (v2.11) as the Monitor and the
Trusted Firmware reference implementation of the RMM~\cite{RMM} (v0.5.0). On
the host, the hypervisor is Linux KVM~\cite{dall2014kvm} from
linux-cca~\cite{linux-cca} (Linux 6.12), and QEMU~\cite{QEMUlinaro} (v10.0)
is our virtual machine manager (VMM) running on the host's userspace. 
Both cVMs run linux-cca (Linux 6.12, v5+v7). The workload cVM runs 
OpenClaw~\cite{OpenClaw} (2026.4.15-beta.1) as the agent, and the
Airlock cVM runs the OpenShell~\cite{OpenShell} supervisor (v0.1.2). 

\para{Extensions} \Cref{tab:extensions} lists the components \name adds
or extends, grouped by function. We modify the RMM, Linux, and
QEMU in C, and write our backends, the policy engine, the ledger, and our OpenShell code in Rust.
Confidential signaling spans three layers. In the RMM, we implement the
interface of \cref{tab:cs-interface}. In KVM, a small change makes the
host use only the LRs the RMM reports. In the guest kernel, a driver maps
each device's doorbells to signaling interrupts.
CIOV has a part shared by both devices and a part per device. In QEMU, we
extend device discovery and plumbing so that each device is exposed over
PCI with its part of the CSM. In the guest kernel, a common transport
places the virtqueues and every buffer in the CSM, so the per-device
changes to the virtio-fs and virtio-vsock frontends are small. In the
Airlock cVM, a shared library maps the CSM and drives the rings and the
signaling for both backends. The CIOV-fs backend is a FUSE server, and
the CIOV-vsock backend hands the rings in the CSM to the kernel's
vhost-vsock.
Policy and audit run in the Airlock cVM. The FS policy engine is part of
the CIOV-fs backend and checks each FUSE request. The agent ledger is a
daemon that chains the file and network records, anchors them in ACL, and
holds the workload's new HTTPS connections until their anchor, through
an egress gate, an HTTP proxy that the OpenShell supervisor uses as its
upstream. In
OpenShell, we add a compute driver that launches the sandbox in the
workload cVM through the launcher, and code that pushes the sandbox's
file system policy to the FS policy engine.

\subsection{Airlock cVM Setup}\label{sec:airlock-setup}
This subsection describes how we set up the Airlock cVM's network and storage on
the performance prototype.

\para{Networking} To enable networking in the Airlock cVM, we rely on the host's
device passthrough capabilities, which give the Airlock cVM direct control of a
network device. We pass the board's physical network interface, an RTL8125
2.5\,GbE controller, to the Airlock cVM with QEMU's
\texttt{vfio-pci}~\cite{vfio,qemupassthrough}. 


\para{Storage} To give the Airlock cVM persistent storage, we dedicate a
partition of the board's eMMC. 
The disk has three partitions. The Airlock cVM boots from the
\textit{init} partition, whose init program mounts the \textit{Airlock cVM}
partition at \texttt{/} as the Airlock cVM's root file system. The Airlock cVM then
mounts the \textit{workload} partition at \texttt{/srv/Airlock} and exports
this directory through the CIOV-fs backend. At boot time, the workload's
kernel mounts CIOV-fs at \texttt{/}.

This network and storage setup minimizes the host's involvement in the
Airlock cVM's I/O, so the I/O overhead we measure for the workload reflects the true
cost of \name rather than that of the host's I/O stack.
Section~\ref{sec:discussion} discusses why this setup is realistic and how it
matches a production confidential deployment.


\section{Evaluation}
\label{sec:evaluation}
We evaluate the performance of \name's confidential signaling and the CIOV devices against their baseline mechanisms (\S\ref{sec:eval-micro}), then measure end-to-end overhead on real agent tasks (\S\ref{sec:eval-macro}).
Since \name introduces no new hardware mechanisms and we observed no stalls, faults, or unexpected behavior throughout our experiments, we expect \name to be compatible with CCA hardware when it becomes available.

\para{Experimental setup}
We run performance experiments on the OpenCCA prototype described in \S\ref{sec:prototypes}, with a total budget of three vCPUs and 3\,GB of memory.
Single-cVM baselines receive the full budget.
With \name, the workload receives two vCPUs and 2\,GB, while the Airlock cVM receives one vCPU and 1\,GB.
The 32\,MB CSM is included in the total memory budget.
We pin all vCPU threads to the Radxa Rock~5B+'s  high-performance Cortex-A76 cores across configurations~\cite{ROCK5Bplus}.

\subsection{Microbenchmarks}\label{sec:eval-micro}

\para{Confidential signaling}
\label{sec:eval-signaling}
\begin{table}[t]
  \centering
  \caption{Round-trip signaling latency between two cVMs over 50,000
    rounds, in $\mu$s.}
  \label{tab:signaling}
  \begin{tabular}{lrrrrr}
    \toprule
    Method & Median & Mean & Std & Min & Max \\
    \midrule
    Confidential signaling & 9.9 & 10.0 & 1.8 & 8.5 & 180.5 \\
    ivshmem signaling      & 38.5 & 38.5 & 1.7 & 31.2 & 238.6 \\
    \bottomrule
  \end{tabular}
\end{table}
%
%
We compare two inter-cVM notification mechanisms: \name's confidential signaling, delivered directly by the RMM without host involvement, and non-confidential signaling through QEMU's ivshmem~\cite{ivshmem}, mediated by the host.
A userspace benchmark exchanges a signal and reply between two cVMs for 50,000 rounds, measuring round-trip latency.

Confidential signaling reduces median round-trip latency from 38.5\,$\mu$s to 9.9\,$\mu$s, a 74.3\% reduction (\cref{tab:signaling}).
It avoids the host exits and re-entries required by ivshmem to forward notifications between the cVMs.


\begin{figure}[t]
  \centering
  \includegraphics[width=\columnwidth]{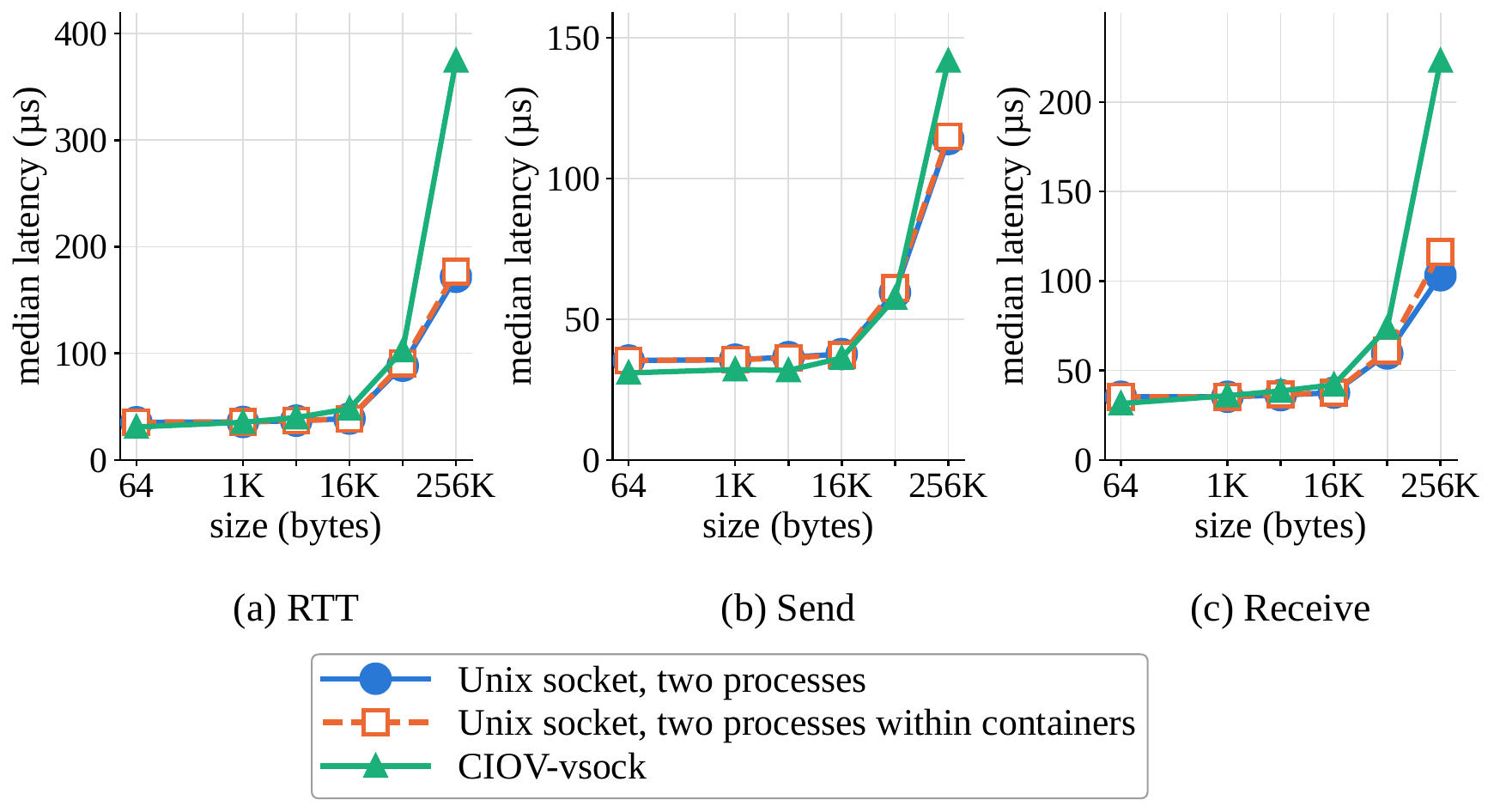}
  \caption{Median socket latency against message size.}
  \label{fig:vsock-lat}
\end{figure}

\begin{figure}[t]
  \centering
  \includegraphics[width=0.9\columnwidth]{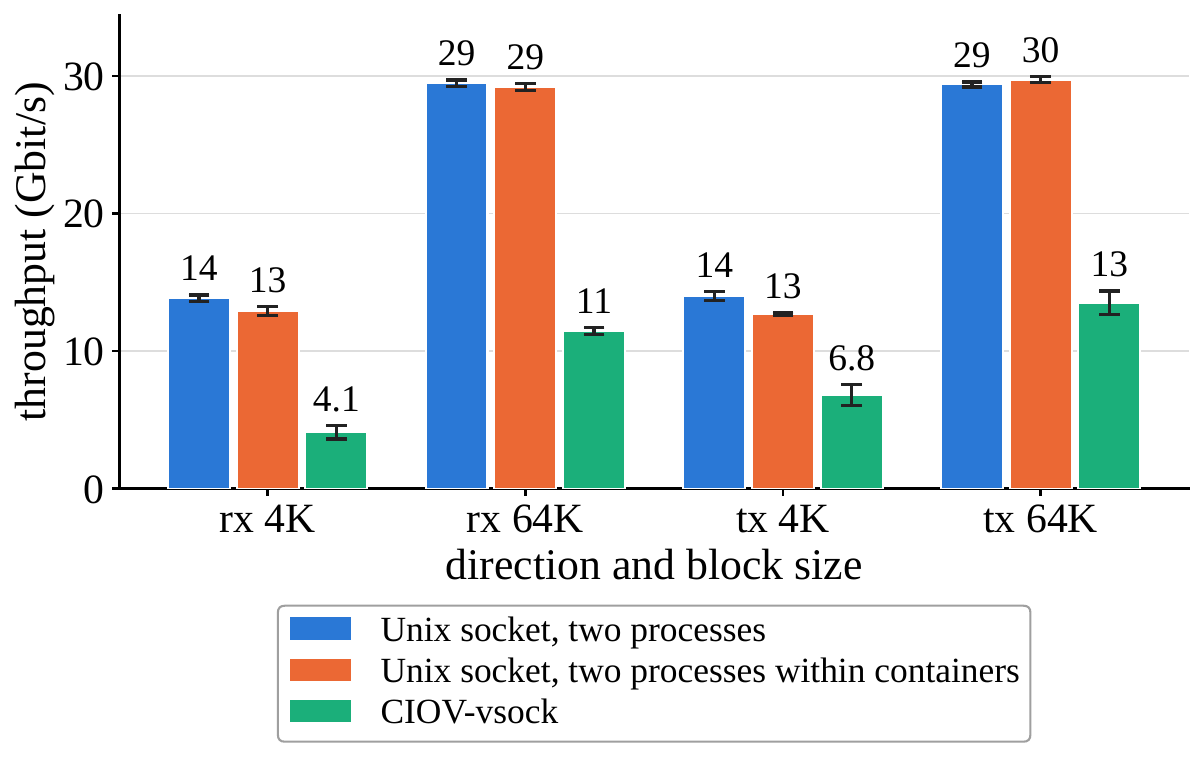}
  \caption{Socket throughput by direction and block size.}
  \label{fig:vsock-tput}
\end{figure}

\para{CIOV-vsock}\label{sec:eval-socket}
We compare three configurations: \textit{baseline}, with two processes communicating over a Unix socket within one cVM; \textit{container}, with the processes in separate containers in the same cVM, matching OpenShell's default broker--supervisor setup; and \textit{\name}, with processes in the workload and Airlock cVMs communicating over CIOV-vsock.
We measure latency for three exchanges: an echo (\textit{RTT}), a variable-size request with a 1-byte reply (\textit{Send}), and a 1-byte request with a variable-size reply (\textit{Receive}).
Message sizes range from 64\,B to 256\,KB, with three runs of 20,000 exchanges per size, reporting median latency.
We also measure throughput in both directions using 4\,KB and 64\,KB blocks, reporting the median of five 10-second transfers.

%
CIOV-vsock reduces latency for short messages but adds overhead for larger transfers (\cref{fig:vsock-lat}).
At 64\,B, its RTT is 30.9\,$\mu$s versus 35.3\,$\mu$s for the baseline, a 12\% reduction.
The configurations converge at 1\,KB; at 256\,KB, CIOV-vsock's RTT reaches 374.5\,$\mu$s versus 171.8\,$\mu$s, a 118\% increase.
Send and receive latencies remain comparable to or below the baseline for messages smaller than 16\,KB.
Throughput ranges from 4.1 to 13.5\,Gbit/s, 51--70\% below the baseline (\cref{fig:vsock-tput}), while the container configuration stays within 13\% of the baseline across all tests.
Overall, CIOV-vsock favors short-message latency over bulk-transfer throughput.

\para{CIOV-fs}\label{sec:eval-fs}
\begin{figure*}[t]
  \centering
  \includegraphics[width=\textwidth]{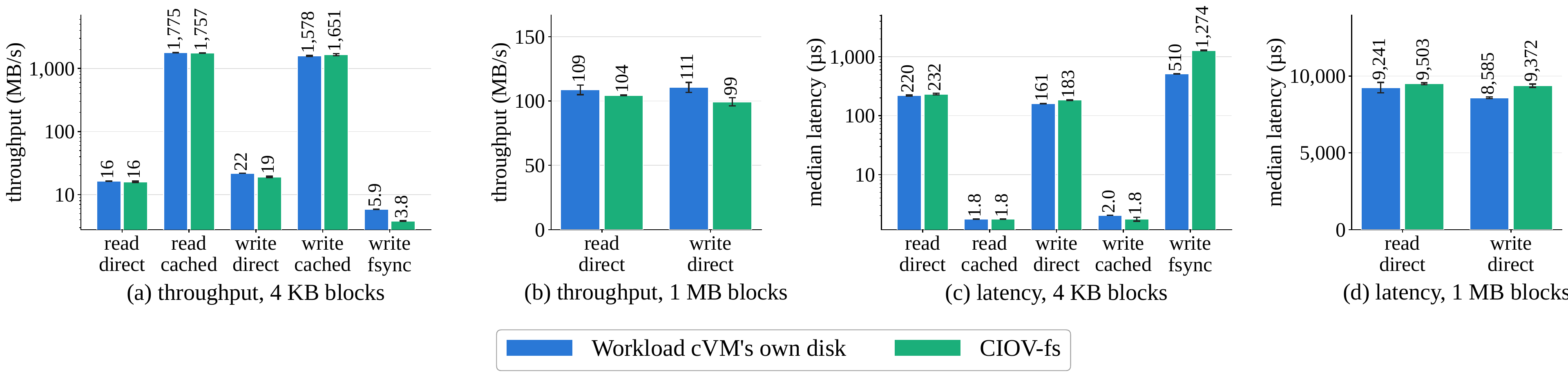}
  \caption{File system throughput (a, b) and median completion latency
    (c, d) of the workload cVM with 4\,KB (a, c) and 1\,MB (b, d) blocks;
    medians of five runs, whiskers one standard deviation, (a) and (c) on a
    log scale.}
  \label{fig:fs}
\end{figure*}
We use fio~\cite{fio} to compare direct disk access from the workload cVM (\textit{baseline}) with access through CIOV-fs, where the Airlock cVM owns the disk (\S\ref{sec:airlock-setup}).
We test three access modes: \textit{direct}, using \texttt{O\_DIRECT} to bypass both cVMs' page caches; \textit{cached}, served from the workload's page cache; and \textit{fsync}, calling \texttt{fsync} after each write to ensure durability.
With 4\,KB blocks at random offsets, we test direct reads and writes (2,000 requests each), cached reads of a preloaded 64\,MB file, cached writes without flushing, and writes followed by \texttt{fsync}.
The cached and fsync tests run for 10 seconds after a 2-second warm-up, with file sizes below the dirty-page limit to avoid writeback throttling.
With 1\,MB blocks, we test sequential direct reads and writes over a 512\,MB file.
We repeat each test five times and report the median.

\Cref{fig:fs} shows throughput and median latency for both block sizes.
Excluding \textit{fsync}, CIOV-fs adds at most 14\% overhead: direct 4\,KB writes incur 14\% higher latency and 13\% lower throughput.
Direct 1\,MB write throughput drops by 10\%, while direct read throughput drops by 3--6\%.
Cached reads remain within 1\% of the baseline, and cached writes show no measurable difference, since both configurations serve these operations mostly from the workload's page cache.

The 4\,KB \textit{fsync} test incurs the largest overhead: throughput drops by 35\% and median latency reaches 2.5 times the baseline.
Each write requires the workload to flush data to the Airlock cVM, which synchronizes the file to disk before replying.
This adds inter-cVM communication and backend processing to every durable write.



\subsection{Macrobenchmarks}\label{sec:eval-macro}
\para{Benchmark setup}
We evaluate end-to-end overhead using OpenClaw~\cite{OpenClaw} in two configurations.
The \textit{baseline} runs unmodified OpenShell in one cVM, including the gateway, supervisor, broker, and agent.
\name separates these components across two cVMs (\S\ref{sec:design}, \S\ref{sec:implementation}).
Both use the same OpenClaw code, sandbox image, policies, and model, GPT-6.1 Sol, accessed through the OpenAI API.
We repeat the \name experiments with and without the agent ledger (\S\ref{sec:ledger}).

\para{Tasks and execution}
We select 12 WildClawBench~\cite{WildClawBench} tasks, grouped by their dominant resource: disk (real\_image, multi\_crop, leaked\_api, risk\_os), memory and compute (jigsaw, chat\_action, product\_poster, file\_overwrite), and network (wiki\_bio, eff\_search, repo\_slides, misinfo).
A Python script follows the benchmark harness through the gateway: staging inputs, initializing the sandbox, submitting the task, and collecting outputs.
Input staging, setup, and output collection are timed separately from task execution.
We repeat each task six times per configuration, issuing an untimed warm-up request before each run.

\begin{table}[t]
  \centering
  \scriptsize
  \setlength{\tabcolsep}{2pt}
  \setlength{\aboverulesep}{0pt}
  \setlength{\belowrulesep}{0pt}
  \renewcommand{\arraystretch}{1.15}
  \caption{WildClawBench per task on the baseline (B) and on \name (A),
    the mean of six runs of each task. Ovh.\ is the increase from B
    to A.}
  \label{tab:macro-time}
  \begin{adjustbox}{max width=\columnwidth}
  \begin{tabular}{l|rr|rrr|rr|rr|rrr}
    \toprule
    & \multicolumn{2}{c|}{Steps} & \multicolumn{3}{c|}{End-to-end (s)} & \multicolumn{2}{c|}{Model (s)} & \multicolumn{2}{c|}{Remote (s)} & \multicolumn{3}{c}{System (s)} \\
    Category & B & A & B & A & \textbf{Ovh.} & B & A & B & A & B & A & \textbf{Ovh.} \\
    \midrule
    Disk    & 8.7  & 7.7  & 66.7  & 64.3  & $-$\textbf{4\%}  & 44.3 & 39.6 & 7.2 & 8.1 & 15.3 & 16.5 & $+$\textbf{8\%}  \\
    Memory  & 10.3 & 10.5 & 86.0  & 95.8  & $+$\textbf{11\%} & 65.8 & 68.1 & 5.5 & 8.1 & 14.7 & 19.6 & $+$\textbf{33\%} \\
    Network & 11.2 & 13.3 & 104.3 & 114.1 & $+$\textbf{9\%}  & 87.6 & 87.8 & 0.2 & 0.3 & 16.5 & 25.9 & $+$\textbf{57\%} \\
    \midrule
    All     & 10.1 & 10.5 & 85.7  & 91.4  & $+$\textbf{7\%}  & 65.9 & 65.2 & 4.3 & 5.5 & 15.5 & 20.7 & $+$\textbf{33\%} \\
    \bottomrule
  \end{tabular}
  \end{adjustbox}
\end{table}

\begin{table}[t]
  \centering
  \scriptsize
  \setlength{\tabcolsep}{3pt}
  \setlength{\aboverulesep}{0pt}
  \setlength{\belowrulesep}{0pt}
  \renewcommand{\arraystretch}{1.15}
  \caption{WildClawBench per model step and per local tool call on the
    baseline (B) and on \name (A), the category's total over its number
    of steps or calls. Ovh.\ is the increase from B to A.}
  \label{tab:macro-step}
  \begin{tabular}{l|rrr|rrr|rrr}
    \toprule
    & \multicolumn{3}{c|}{Model (s/step)} & \multicolumn{3}{c|}{Local (s/call)} & \multicolumn{3}{c}{System (s/step)} \\
    Category & B & A & \textbf{Ovh.} & B & A & \textbf{Ovh.} & B & A & \textbf{Ovh.} \\
    \midrule
    Disk    & 5.11 & 5.17 & $+$\textbf{1\%}  & 1.25 & 1.25 & \textbf{0\%}     & 1.76 & 2.15 & $+$\textbf{22\%} \\
    Memory  & 6.36 & 6.46 & $+$\textbf{2\%}  & 0.72 & 0.86 & $+$\textbf{20\%} & 1.43 & 1.86 & $+$\textbf{31\%} \\
    Network & 7.81 & 6.59 & $-$\textbf{16\%} & 1.18 & 1.63 & $+$\textbf{39\%} & 1.47 & 1.95 & $+$\textbf{32\%} \\
    \midrule
    All     & 6.54 & 6.20 & $-$\textbf{5\%}  & 1.04 & 1.28 & $+$\textbf{23\%} & 1.54 & 1.97 & $+$\textbf{28\%} \\
    \bottomrule
  \end{tabular}
\end{table}

\para{Metrics} We measure \textit{end-to-end} latency from task submission to OpenClaw's response and split it into three components.
\textit{Model time} covers model API calls; \textit{remote-tool time} covers image analysis, web search, web fetch, and browsing.
The remaining \textit{system time} includes local shell commands and file accesses, request handling, and OpenClaw's processing between model calls.
We use system time to assess \name's local overhead, excluding model and remote-service delays, while end-to-end latency captures the user-visible impact.
Each model call counts as one \textit{step}; since the agent's execution path varies across runs, we report times per task (\cref{tab:macro-time}), per step, and per local tool call (\cref{tab:macro-step}).
Values are means over six runs per task, and \textit{Ovh.}\ denotes the relative increase with \name.
Both tables show \name without the agent ledger.
We do not assess output correctness.

\para{Per task}
\name increases mean system time per task from 15.5 to 20.7\,s (33\%), while mean end-to-end latency rises from 85.7 to 91.4\,s (7\%) (\cref{tab:macro-time}).
The user-visible overhead is smaller because model calls dominate execution, accounting for over 70\% of end-to-end latency in both configurations.
Mean model time is 65.9\,s in the baseline and 65.2\,s with \name, while remote tools account for 4.3\,s and 5.5\,s, respectively.

\para{Per step}
As the number of steps varies across runs and directly impacts end-to-end execution time, we also normalize times by the number of model steps or local tool calls (\cref{tab:macro-step}).
System time per step increases from 1.54 to 1.97\,s (28\%), with increases of 22--32\% across task categories.
Model time per step decreases from 6.54 to 6.20\,s (5\%), reflecting variation in model calls rather than local system performance.

\para{Per local tool call}
Mean local tool latency rises from 1.04 to 1.28\,s per call (23\%): unchanged for disk-heavy tasks, 20\% higher for memory-heavy tasks, and 39\% higher for network-heavy tasks.
The negligible overhead for disk-heavy tasks is consistent with the microbenchmarks' near-baseline cached file-system performance, while CIOV-vsock's lower bulk-transfer throughput may contribute to the higher overhead on network-heavy tasks.


\para{Agent ledger overhead}
We repeat the benchmark on \name with the agent ledger enabled and anchoring its records in ACL.
Most of the ledger's cost comes from network connections.
Each new HTTPS connection waits at the egress gate until its record is anchored in ACL.
This wait adds 259\,ms per connection on average, most of which is ACL's median response time of 213\,ms.
A task opens 14.6 such connections on average, mostly for model calls, so the gate delays each task by 3.8\,s in total.
File accesses, in contrast, do not wait for ACL.
The CIOV-fs backend serves each file request at once and only queues a record of it for the ledger.
This takes 0.7 to 1.3\,$\mu$s per file request, or 0.07\,s per task on average.
Even the most file-intensive run, real\_image with 236k file requests, spends only 0.16\,s on it.
Overall, mean end-to-end latency rises from 91.4 to 95.0\,s (4\%) over \name without the ledger.

\section{Discussion}
\label{sec:discussion}
%
In this section, we discuss gaps between our prototype and a real-world
deployment of \name, and possible future research directions and extensions.

\para{Network passthrough} Relying on passthrough does not limit \name's
practicality in real-world deployments. Modern network interface cards
(NICs) let the host assign a VM an entire PCIe \textit{physical function}
(PF), as in our prototype, or, with SR-IOV~\cite{sriov}, a lightweight
\textit{virtual function} (VF), keeping the host off the data path.
Confidential computing platforms extend this assignment: a cVM attests an
assigned function before use, and the path between them is protected from
the host~\cite{linuxtsm}. Arm CCA supports this through RME device
assignment (RME-DA)~\cite{rmeda}, so \name's networking carries over to
production hardware without change.

\para{Disk protection} Fully protecting the state of a cVM requires
extending protection beyond its memory to its disk. Existing systems
protect the disk's confidentiality and integrity~\cite{dmcrypt,dmintegrity},
and ShieldFS~\cite{shieldfs} additionally detects rollback, keeping the
disk's state fresh. We prototyped \name's integration with ShieldFS and
report our initial design and experience in \S\ref{app:shieldfs}.

\para{Performance overhead} We expect \name's overhead to be lower on CCA
hardware, mainly because of device interrupts. \name's I/O paths complete
through such interrupts: the Airlock cVM's NIC raises one for each batch of
packets it receives or sends, and the Airlock cVM's disk raises one for each
completed request. On our board, whose interrupt controller implements
GICv3, every such interrupt exits the Airlock cVM to the host, which injects it
and resumes the Airlock cVM, crossing the RMM and the EL3 monitor twice. GICv4
removes this round trip with \textit{direct injection}: the interrupt
controller delivers a device's interrupt straight to the running vCPU,
much like posted interrupts on x86~\cite{armgicv4spec,armgicvirtualization}.
Once CCA supports direct injection for cVMs, the Airlock cVM could receive
many of its network and disk completions without exits to the host,
similar to how \name's inter-cVM notifications avoid the host through
confidential signaling (\S\ref{sec:conf-signaling}).

\para{GICv5} 
Confidential signaling wakes a running peer with an SGI. GICv3 has no
interrupt group for the Realm world and treats the Realm world's SGIs as
Non-secure interrupts~\cite{armgicv4spec}. The state of this SGI lives
in the target core's redistributor, and since the host can access the
Non-secure state of the redistributor through MMIO, it can observe the
SGI state. GICv5 removes this exposure.
Its Interrupt Routing Service (IRS) replaces the Distributor and the
redistributors and keeps a separate interrupt domain for the Realm
world, whose registers and interrupt state are out of the host's
reach~\cite{gicv5}.
The RMM could then wake a peer with a Realm-domain interrupt that the
host can neither read nor set. To hide its signals from the host, \name
therefore relies on GICv5, which we expect future CCA hardware to
provide.

\para{Other platforms}
Extending \name to Intel TDX~\cite{inteltdx} and AMD SEV-SNP~\cite{amdsevsnp}
requires platform-specific mechanisms for protected I/O buffers, authenticated
notifications, and attestation that binds the workload to its enforcement domain.
Access to shared memory encryption keys and signaling can be implemented
in the TDX Module to implement CIOV.
However, AMD SEV-SNP is more difficult to support. An alternative is to
implement the Airlock cVM's role in a nested VM using VMPL or TD partitioning~\cite{intel2023tdpartition}.

\para{Offloading the Airlock cVM to a DPU}
A second direction is to offload the Airlock cVM's mediation function to a
DPU supporting the TEE Device Interface Security Protocol
(TDISP)~\cite{pcisigtdisp}. Network and file-system policies would execute
directly on the DPU, with credentials and audit generation outside the
workload and host. TDISP, together with device attestation and protected
interconnects, provides a foundation for trusted device assignment.
The design must also protect DMA buffers and bind the assigned interface
and DPU policy state to the workload, leaving no independent path around
enforcement. This direction is promoted by Nvidia in its Sentry~\cite{arkin2026sentry}
architecture for the BlueField-4 DPU.

\section{Related Work}
\label{sec:related}
\para{Confidential VM compartmentalization}
Hecate~\cite{ge2022hecate} repurposes SEV-SNP's Virtual Machine Privilege
Levels (VMPLs) to run an unmodified nested guest under a trusted in-cVM
hypervisor, without relying on the outer hypervisor for nested
virtualization. It already supports policy enforcement, including network
firewalls and kernel-integrity checks. NestedSGX~\cite{wang2025nestedsgx}
uses VMPLs to provide SGX-compatible enclaves protected from an untrusted
guest OS. VeriSMo~\cite{zhou2024verismo} verifies an SEV-SNP security module
providing code integrity, runtime measurement, and secret management.
Intel's TD partitioning specification~\cite{intel2023tdpartition} likewise
supports nested VMs within a TDX trust domain under a trusted L1 VMM.
\name instead separates the workload and supervisor into peer cVMs and
connects them through confidential virtual I/O devices, with policy and
audit generation outside the workload's compromise domain.

\para{Protected communication and attestation}
CAEC~\cite{abdollahi2025confidential} introduces confidential shared memory
between CCA realms, avoiding encryption over host-accessible buffers.
MAGE~\cite{chen2022mage} addresses mutual attestation of enclave groups.
MICA~\cite{mica} extends CCA firmware with restrictions on communication
channels and group attestation. \name extends these works with confidential
signaling and CIOV.
AgenTEE~\cite{abdollahi2026agentee} separates the agent runtime, inference
engine, and third-party components into independently attested cVMs, but
does not consider malicious agents.

\para{Runtime privilege control}
Progent~\cite{shi2025progent} checks tool names and arguments against symbolic
policies. Its updated design permits automatic privilege narrowing but
requires approval for expansions. AgentSpec~\cite{wang2026agentspec}
expresses runtime constraints as triggers, predicates, and enforcement
actions. AgentBound~\cite{buhler2026agentbound} applies declarative
resource permissions to unmodified MCP servers using container-based
enforcement, trusting Docker and the host kernel.
\name addresses the placement of enforcement when the workload kernel
and host are malicious, and composes with tool-calling policies.

\para{Information-flow control}
Decentralized information-flow control~\cite{myers1997ifc} and
HiStar~\cite{zeldovich2006histar} provide foundations for enforcing
data-dependent policies; HiStar places enforcement in a trusted kernel.
CaMeL~\cite{debenedetti2026camel} separates trusted control flow from
untrusted tool results and checks capabilities before tool calls to
restrict unauthorized data flows. Fides~\cite{costa2025fides}
tracks confidentiality and integrity labels across agent interactions
and selectively hides data to preserve security while supporting useful plans.
The design-pattern analysis of Beurer-Kellner et al.~\cite{beurer2025patterns}
places these mechanisms in a broader space of constrained agent architectures.
Their guarantees depend on the trusted runtime mediating the relevant
actions and thus are complementary to \name: information-flow
policies can reject exfiltration through otherwise permitted tool calls,
whereas we enforce file and network policies without requiring
agent-wide label tracking.

\para{Prompt injection and model-level defenses}
Greshake et al.~\cite{greshake2023not} demonstrate indirect prompt injection
against applications that combine LLMs with external data. InjecAgent~\cite{zhan2024injecagent}
and AgentDojo~\cite{debenedetti2024agentdojo} provide benchmarks for attacks
and defenses involving tool-integrated agents. StruQ~\cite{chen2025struq}
separates instructions from data through structured queries and training,
while SecAlign~\cite{chen2025secalign} uses preference optimization to improve
resistance to injected instructions. These approaches address how models
interpret untrusted input. \name instead treats the workload as potentially
malicious, including its native code and guest kernel, and protects
enforcement independently of model behavior.

\para{Persistent state and verifiable auditing}
Disk encryption~\cite{dmcrypt} does not establish freshness against rollback.
ShieldFS~\cite{shieldfs} protects persistent file-system integrity and
freshness despite an untrusted storage stack. This protection is a separate
layer from CIOV's live communication guarantees.
CCF~\cite{russinovich2019ccf} and Azure Confidential Ledger~\cite{azureacl}
provide verifiable ledger infrastructure. We build on these primitives to
implement CIOV-fs and the agent ledger.

\section{Conclusion}
\label{sec:conclusion}
We presented \name, an architecture that places policy enforcement and
audit generation in an Airlock cVM outside the workload's compromise
domain. Its confidential I/O virtualization combines protected shared
memory, firmware-level signaling, and socket and file-system devices whose
backends run in the Airlock cVM, preserving mediation even after workload-kernel
compromise under the stated trust assumptions.
Our architecture applies to local and hosted agents on CCA-capable platforms
such as Nvidia's Vera or Microsoft's Cobalt CPUs.
Our architecture achieves small I/O overheads and is designed to integrate
with many agent platforms beyond OpenShell.

\ifCLASSOPTIONcompsoc
  \section*{Acknowledgments}
  The authors have used generative AI (Claude Opus 5.5 and Fable 5.1) as auxiliary tools for editorial purposes,
finding related research, and code debugging. 
All generated material was examined and verified by the authors, 
who take full responsibility for the accuracy, integrity, and originality of the paper and the released code.

\else
  \section*{Acknowledgment}
    The authors have used generative AI (Claude Opus 5.5 and Fable 5.1) as auxiliary tools for editorial purposes,
finding related research, and code debugging. 
All generated material was examined and verified by the authors, 
who take full responsibility for the accuracy, integrity, and originality of the paper and the released code.

\fi


\bibliographystyle{IEEEtran}
\bibliography{bib}

\appendices
\crefalias{section}{appendix}
\section{CIOV-vsock Details}\label{app:ciov-socket}
This appendix details the two performance optimizations of CIOV-vsock
that \S\ref{sec:ciov} outlines: the sizing of receive buffers, the one
change \name makes to stock virtio-vsock, and the in-kernel backend that
serves the socket in the Airlock cVM.

\para{Receive buffers} Receive buffers are those the workload posts for
data coming from the Airlock cVM. Since the driver must post them before it
knows what will arrive, it has to choose their size in advance, and stock
virtio-vsock always chooses 4\,KB. A 64\,KB message therefore costs
sixteen slots, sixteen copies, and sixteen wake-ups.

\name instead borrows the mergeable receive buffers of
virtio-net~\cite{virtiospec}: the driver keeps a running average of the
sizes of the packets it has received and posts each new buffer at that
average, between 4 and 64\,KB, so that a stream of large messages soon gets
large buffers and a stream of small ones does not waste pool space. In our
evaluation this doubled receive throughput without changing small-message
latency (\S\ref{sec:eval-socket}).

The transmit path needs no such change, since the driver knows the size of
the data it is about to send and sizes each buffer to it, up to the
protocol's 64\,KB maximum.

\para{In-kernel backend} The CIOV-vsock backend in the Airlock cVM can serve
the virtqueues itself, in user space, or hand the data processing to the
Airlock cVM's kernel. For the latter it reuses \textit{vhost-vsock}~\cite{vhostvsock},
Linux's in-kernel virtio-vsock backend: a kernel thread that serves a device's
virtqueues directly, so that no packet passes through a user-space process
and each one is copied once, from the virtqueue's buffer into the kernel's
socket. \name's backend sets up vhost-vsock and then steps aside. It maps
the CSM and installs it, together with the virtqueues, in vhost. It then
bridges \name's signals to the notification mechanism vhost expects: when
the workload signals, the backend wakes vhost's thread, and when vhost has
completed work, the backend signals the workload. After this
setup the backend becomes idle. The kernel moves the data, and the socket
stream it rebuilds is delivered to the Airlock cVM's socket layer. We use this
mode by default, since it keeps a process off the data path and
outperformed the user-space mode on our board.

\section{File Reference Monitor Details}\label{app:frm}
This appendix details the file reference monitor that \S\ref{sec:overview}
introduces.

The \textit{backend} is the transport. It takes every FUSE request the
workload issues, whether it comes from the agent or from any
other process in the workload, hands it to the FS policy engine, and performs
it only if the engine allows it.
The backend holds no policy and makes no decision.
The uid, gid, and pid in a FUSE request's header are
claims the workload's kernel makes about the caller; the engine interprets
them.

The \textit{FS policy engine} decides each request. A policy is a list of
path rules, each granting read or write access to a path and, for a
directory, to everything beneath it.

The engine holds a stack of layers, each a Landlock ruleset that names the
uid and gid it governs, and allows a request only if every layer that
names its sender allows it. Boot layers, which \name ships with the
Airlock cVM and the engine loads from its disk before the workload boots,
govern the workload's root processes and give the sandbox's user its base
view. When the supervisor starts a
sandbox, it pushes the user's file-system policy to the engine as further
layers scoped to the sandbox's user: the same rulesets the broker
compiles into the agent's Landlock ruleset, so one policy drives both
layers. Each pushed layer can only narrow the boot layers, and a request
from a user that no layer names is refused. The engine fixes a
file's rights at open and remembers them per handle.
It emits one record per decision to the ledger, whether it allowed
or denied the request, and one record for every policy it loads or
removes, with the policy's hash and its origin, so the ledger shows which
policy was in force for each decision.

\section{Confidential Signaling Authorization}\label{app:cs-auth}
CS reuses MICA to decide who may signal whom (\S\ref{sec:conf-signaling}).
A MICA policy can carry a signaling object (\texttt{Sig}) next to its
shared-memory object (\texttt{Mem}). The signaling object refers to the
peers the MICA policy already identifies, and gives each peer a range of
interrupt IDs, called \textit{lanes}, that it may receive. In the example below, \texttt{VM1} and
\texttt{VM2} are the workload and the Airlock cVM, and each receives nine
lanes, at interrupt IDs 208--216.\par\smallskip
\noindent\begin{minipage}[t]{\columnwidth}
\footnotesize
\begin{verbatim}
  "Sig1": {
    "VM1": {"rx-intid": 208, "rx-lanes": 9},
    "VM2": {"rx-intid": 208, "rx-lanes": 9}
  }
\end{verbatim}
\end{minipage}\par\smallskip
\noindent Lanes 0--7 are a fixed pool reserved for the CIOV devices. Each
virtqueue has its own lane, two for CIOV-fs and three for CIOV-vsock, and
the rest of the pool stays free. Lane~7 is a control lane: when the
workload resets a device, it signals lane~7 so that the device's backend in
the Airlock cVM wakes up and stops. Lane~8 serves a user-space signaling
device that our microbenchmark uses (\S\ref{sec:eval-signaling}). A range
starts at its first interrupt ID, so the nine lanes cover the whole pool
and lane~8.

As with shared memory, the RMM activates a signaling link only if the
MICA policies of both peers declare the same link (\S\ref{sec:attestation}).
A sender names only a lane. The RMM finds the caller's peer from the
caller's identity, checks the lane against the peer's range, and makes the
matching interrupt pending, \eg lane~0 of the workload raises
interrupt~208 in the Airlock cVM. Once it accepts a cVM's MICA policy, the
RMM also refuses any interrupt in that cVM's lanes that the host tries to
inject, so the host cannot forge a signal. Finally, the signaling object is
part of the MICA policy, so the RMM extends it into the cVM's REM with the
rest of the MICA policy, and the attestation token reports the signaling
links as well.

\section{Agent Ledger Details}\label{app:ledger}
This appendix details two mechanisms of the agent ledger that
\S\ref{sec:ledger} outlines: how the ledger chains its records and anchors
them in ACL, and which of the agent's actions wait for the anchor.

\para{Chaining and anchoring} The ledger serializes each record
canonically, numbers it, and chains it to the previous record by that
record's SHA-256 hash. For each window, it computes a Merkle root over
the window's records and writes one entry to ACL: the sandbox's
identity, the Airlock cVM's boot, the window's sequence range, its
root, and the chain head. Each sandbox and boot gets its own ACL
collection. ACL returns a signed
receipt for each entry, which the ledger stores next to the window. The
entry also carries the window's records themselves, in order, as many as
fit within ACL's limit of 8\,KiB per entry, and counts per source of the
records left out. When all of a window's records fit, its root can be
recomputed from ACL alone. One entry per window, however many records the window holds, keeps
the cost of ACL to one round trip per window rather than one per action.
Several windows can be in flight to ACL at once, so a window does not
wait for the previous one to commit. When nothing happens, the ledger
still writes a heartbeat record every ten seconds, so a verifier can
tell an idle agent from a ledger that stopped anchoring.

\para{Synchronous and asynchronous recording} The ledger records the
agent's network and file actions differently. New HTTPS connections are
recorded \textit{synchronously}. A connection cannot be undone once it
has left the device, so the egress gate opens a new HTTPS connection
only once the window holding its record is anchored in ACL, and the
ledger closes that window at once rather than at the end of its period.
Plain HTTP requests and DNS queries are recorded as they leave, without
waiting for the anchor, and inbound traffic is not recorded. File accesses are
recorded \textit{asynchronously}. The Airlock cVM serves each request at
once and records it as it happens, and the record is anchored with its
window, every two seconds by default, without the request waiting for
ACL.

\section{Disk Protection with ShieldFS}\label{app:shieldfs}
This appendix details the integration of \name with
ShieldFS~\cite{shieldfs} that \S\ref{sec:discussion} outlines.

ShieldFS creates an encrypted storage \textit{pool} on a disk partition
and holds one or more file systems, called \textit{datasets}, in that
pool. On every sync, it computes a commitment to the pool's new state and
waits until an external registry acknowledges it, so that a rolled-back
disk is detected when the pool is opened. The current implementation,
however, tracks commitments for only one pool per cVM.

Integrating ShieldFS therefore required two changes to our prototype.
First, the Airlock cVM's and the workload's root file systems can no longer
occupy two separate partitions (\S\ref{sec:airlock-setup}), since each
partition would need its own pool. We thus merge the two partitions into
a single partition that holds one pool, and store each root file system
in it as a separate dataset; the Airlock cVM opens the pool at boot. Second,
the Airlock cVM uses ACL as ShieldFS's registry, in a new collection alongside
that of the agent's ledger.

We repeat the experiment of \S\ref{sec:eval-fs} on this pool. In the
baseline, the workload cVM opens the pool itself as its own disk and
registers the commitments in ACL. In \name's mode, the Airlock cVM opens the
pool, registers the commitments, and serves the workload's dataset over
CIOV-fs. Both modes thus pay for the same disk protection
and differ only in \name's file system virtualization. 

Two results stand
out. First, \name's overhead over the baseline remains small: reads and
direct writes stay within 13\% of the baseline (\eg 14.2 against
13.9\,MB/s for 1\,MB direct reads), with buffered writes again the
exception (\S\ref{sec:eval-fs}). Second, the protection itself costs
roughly an order of magnitude in both modes (\eg 1\,MB direct reads drop
from 112 to 14\,MB/s in the baseline). This cost stems from our
prototype's hardware, not from ShieldFS's design: the RK3588 has neither
Arm's cryptographic extensions nor a cryptographic accelerator, so every
cipher and hash runs in software. Current Arm cores implement these
extensions, so we expect the cost of disk protection to be far lower on
production hardware.

%




\end{document}